\documentclass[preprint,useAMS,usegraphicx,usenatbib]{mn2e}

\title[Localized enhancements at oblique shocks]{Localized enhancements of energetic particles at oblique collisionless shocks}
\author[F. Fraschetti and J. Giacalone]{F. Fraschetti\thanks{E-mail:
ffrasche@lpl.arizona.edu (FF)} and J. Giacalone\\
Departments of Planetary Sciences and Astronomy, University of Arizona, Tucson, AZ, 85721, USA}

\usepackage[usenames,dvipsnames]{xcolor}
\usepackage{color}
\usepackage{amssymb,amsmath}
\usepackage{epstopdf}
\usepackage[colorlinks,urlcolor=blue,citecolor=blue,linkcolor=blue]{hyperref}
\usepackage{marginnote}

\newcommand{\vect}[1]{\mathbf{#1}}   % Vector as bold

\begin{document}

\date{Accepted .... Received ...; in original form ...}

\pagerange{\pageref{firstpage}--\pageref{lastpage}} \pubyear{2014}

\maketitle

\label{firstpage}

\begin{abstract}
We investigate the spatial distribution of charged particles accelerated by non-relativistic oblique fast collisionless shocks using three-dimensional test-particle simulations. We find that the density of low-energy particles exhibit a localised enhancement at the shock, resembling the ``spike'' measured at interplanetary shocks. In contrast to previous results based on numerical solutions to the focused transport equation, we find a shock spike for any magnetic obliquity, from quasi-perpendicular to parallel. We compare the pitch-angle distribution with respect to the local magnetic field and the momentum distribution far downstream and very near the shock within the spike; our findings are compatible with predictions from the scatter-free shock drift acceleration (SDA) limit in these regions. The enhancement of low-energy particles measured by {\it Voyager 1} at solar termination shock is comparable with our profiles. Our simulations allow for predictions of supra-thermal protons at interplanetary shocks within ten solar radii to be tested by Solar Probe Mission. They also have implications for the interpretation of ions accelerated at supernova remnant shocks.
\end{abstract}


\section{Introduction}

Collisionless shocks are considered major sources of energetic particles in 
interplanetary and interstellar medium. The observational association of
energetic particles with shocks ranges from (keV-MeV) for solar energetic particles 
originating in solar events \citep{lario03,g12} to the galactic cosmic rays 
whose TeV signature has been observed in extended shell-type supernova remnants \citep{Magic07,a11}.

Localised enhancements of energetic particles are often seen at the time of passage of interplanetary shocks by energetic particles detectors. As recently reported by \citet{lario03}, in addition to the classic energetic storm particle events (a few hours), structures seen at travelling interplanetary shocks, e.g., by {\it ACE} energetic particle detector, also spike events ($\sim 10$ min or shorter) are observed. Enhancements of energetic proton intensities at shocks were observed by Explorer 33-35 \citep{a70} or {\it Vela 4} \citep{sm71}. 
\citet{ar63} interpreted these as evidence of anisotropic reflection of energetic particles between Earth's bow shock and shocks travelling in the solar wind. This interpretation was questioned by \citet{a70} who found in some events energetic-particle features behind the shock similar to the upstream. Spike events were modelled by \citet{d83} who considered shock drift acceleration (SDA) 
of a seed particle population at a quasi-perpendicular shocks ($\theta_{Bn} > 70^{\rm o}$, 
where $\theta_{Bn}$ is the angle between the shock normal and the upstream ordered magnetic field).
However, an extension of the SDA scenario in \citet{d83} was needed to explain the energetic storm particle events, or more generic post-shock enhancements, associated with $\theta_{Bn} < 70^{\rm o}$. \citet{s85} argued  that the self-excited waves produced upstream by energetic ions via streaming instability might account for the local enhancements at shocks with $\theta_{Bn} < 70^{\rm o}$. Upstream magnetic traps for energetic particles formed by multiple encounters of the field lines with the shock surface have also been invoked as a way to generate spikes through the collapse of the trap into the shock as the field advects with the fluid \citep{eb94}. The conditions on the shock obliquity under which particle anisotropy leads to shock spike were analysed in \citet{gkga99}. The crossing of the solar termination shock by {\it Voyager-1} \citep{d05} also seemed to exhibit a spike in high-energy protons ($3.4 -17.6$ MeV) and ions ($40 -53$ keV) intensities, as  was suggested by \citet{l07} and \cite{zf13}.

The spatial scale of the increase of particle intensity that we refer to as a ``spike'' is of the order of gyroscale, suggesting that the diffusion approximation does not apply to the spike formation. Thus, there must exist an anisotropy in the pitch-angle distribution. The origin of energetic particles intensity spikes across collisionless oblique non-relativistic shocks have been considered in several recent studies, using two different approaches with diverging outcomes: 1) Monte-Carlo simulations \citep{ebj95}, i.e.,
the trajectories of a large number of particles are simulated with an ad hoc prescription 
for the pitch-angle scattering \citep{ebj96} with no emerging spike; 2) focused transport equation, or FTE, \citep{i97,kj04} which relaxes the assumption of near isotropy in the pitch-angle distribution and does not assume continuity of the particle distribution across shocks (see, e.g. \cite{l07}). 
Contradictory results about the formation of shock spike were reported also  
for relativistic shocks, as shown by Monte Carlo test-particle simulations \citep{o91,nt95}.

Observational consequences of spikes are relevant also to astrophysical shocks. 
Piling-up of energetic electrons at the shock might result in an enhanced filamentary 
synchrotron-emitting structures at supernova remnant shocks, as argued in \citet{gkga99}. 
An enhancement of particle density at the shock might also affect filamentation structures 
thought to be seeded by Weibel instability at relativistic shocks \citep{ml99}.

The spike-like discontinuity found in the upstream particle density was interpreted as due to particles 
reflected after one shock-crossing and conserving the magnetic moment
as they encounter the magnetic barrier at quasi-perpendicular shocks \citep{gkga99}. 
However, for smaller obliquity shocks this interpretation does not seem to apply.
Previous approaches do not clarify the dependence of the spike intensity on $\theta_{Bn}$. 
A shock spike emerges from the solution to the FTE regardless of $\theta_{Bn}$; 
however, for quasi-parallel shocks it has been argued \citep{l07}
that the spike is unphysical and it should disappear if the particle momentum were calculated in the shock rest frame, 
instead of being calculated in the local plasma frame, as in the \cite{p65} approximation.

In this paper we present results from the numerical integration of test-particle trajectories in kinematically prescribed turbulent fields associated with a collisionless shock. The geometry comprises a planar shock with an arbitrary obliquity of the upstream average field. Particle scattering and decorrelation arises from directly integrating the particle equation of motion in the given turbulent background, without {\it ad hoc} prescription of pitch-angle change. This procedure provides us with spatial profiles of the intensity of energetic particles with an unexpectedly significant enhancement at quasi-parallel shock. We compare the pitch-angle anisotropy and the momentum distribution at the shock with far downstream. Our density profiles are contrasted with the DSA prediction in various particle momentum range, from supra-thermal up to energetic particles. As in our previous works \citep{fg12}, 
parallel and perpendicular diffusion originate from a fully specified magnetic turbulence.

This paper is organized as follows: in Sect. \ref {Non-diff} previous extensions to non-diffusive regime 
are outlined; in Sect. \ref{numerical}, complemented by the appendix, 
we describe details of the numerical approach. In Sect. \ref{trajectories} we present sample charged particle 
trajectories at the shock, identifying the role of magnetic obliquity and turbulence power 
in the energization process. In Sect. \ref{profiles} we show that spatial density profiles
exhibit an enhancement at the shock regardless of the shock obliquity and for sufficiently low-energy particles. 
In Sect. \ref{anisotropy} we compare the distribution 
of the pitch-angle with respect to the local fluctuating field
ahead and behind the shock with the expected {\it near}-isotropy assumed by DSA and with the model
for scatter-free SDA. In Sect. \ref{spectra} we compare the momentum spectrum at the spike with the far downstream high-momentum power-law. In Sect. \ref{helio} we discuss the implications of our findings for the {\it in-situ} measurements of interplanetary shocks.
In Sect. \ref{discussion} we discuss the findings and outline the conclusions.

\section{Non-diffusive approach}\label{Non-diff}

Standard DSA theory associates to shocks a population of energetic particles 
with a pitch-angle distribution {\it nearly} isotropic in the local plasma rest frame.
This implicitly assumes that the speed $v$ of those particles
in the local plasma frame is much greater than the speed of the shock in the upstream fluid $U_1$.
In other words particles have time to scatter on the magnetic irregularities and 
isotropize their pitch-angle distribution as the bulk fluid advection occurs on much greater time-scale.
In the limit that particles adhere to magnetic field lines, 
$v$ is to be compared with the shock speed in the upstream fluid 
projected along the upstream average field, i.e. $U_1/ cos \theta_{Bn}$,
namely the speed along the shock of the intersection point field line/shock. 
Clearly this ought not to be valid if decorrelation from field line is allowed.

The steady-state solution of the Parker transport equation \citep{p58} for a one-dimensional planar shock and uniform plasma flows on either side predicts that the particle density rises exponentially 
from far upstream toward the shock, with an {\it e}-folding distance depending 
on the spatial coefficient diffusion $\kappa$ in the direction perpendicular 
to the local shock surface and on the fluid advection speed $U_1$. 
The intensity is continuous at the shock along with its Lagrangian derivative 
and in the downstream flow it is uniformly equal to a constant depending only
on the particle momentum. 
In the presence of self-generated waves the steady state upstream profile rarefies more slowly than exponentially  ($1/x$) \citep{b78}.

Deviations from the assumed {\it near}-isotropy have been already studied in several cases 
even in the absence of shocks. 
For instance, scattering on small-scale turbulence is found to generate an inertial force due to the velocity shear of second order in $U_1/v$ \citep{ejm88}, usually neglected with respect to 
the adiabatic energy change in the transport equation ($\nabla \cdot {\bf U}$).
If the random motion of scattering centres is embodied in the momentum diffusion coefficient,
the standard second-order in $U_1/v$ Fermi acceleration arises. 
Neglecting orders higher than first in $U_1/v$ requires disregarding 
the anisotropic part, in the local plasma frame, of the phase-space particle distribution function,
although one can solve for the isotropic part of $f$ including second order momentum diffusion.

\section{Numerical set-up}\label{numerical}

We carry out numerical simulations of supra-thermal \footnote{Here supra-thermal particles 
have speed $v$ greater but comparable with the upstream flow speed in the shock frame, $U_1$, in contrast with high-energy particles ($v \gg U_1 $).} test-particles (protons) 
at collisionless fast shocks travelling in a medium with an embedded turbulent magnetic field
with a moderately high upstream Alfv\'en Mach number, i.e., $M_A \le 10$ (compression of transverse components of ${\bf U}$ and ${\bf B}$ depends on $M_A$). We consider the limit of the initial particle speed  comparable to the bulk plasma flow speed. In our model, particle scattering arises from the interaction between the particles and a kinematically prescribed turbulent magnetic field. We examine steady-state particle density profiles across the shock for various particle energies, variance in the turbulent random magnetic field, and $\theta_{Bn}$.

\begin{figure}
\includegraphics[width=10cm]{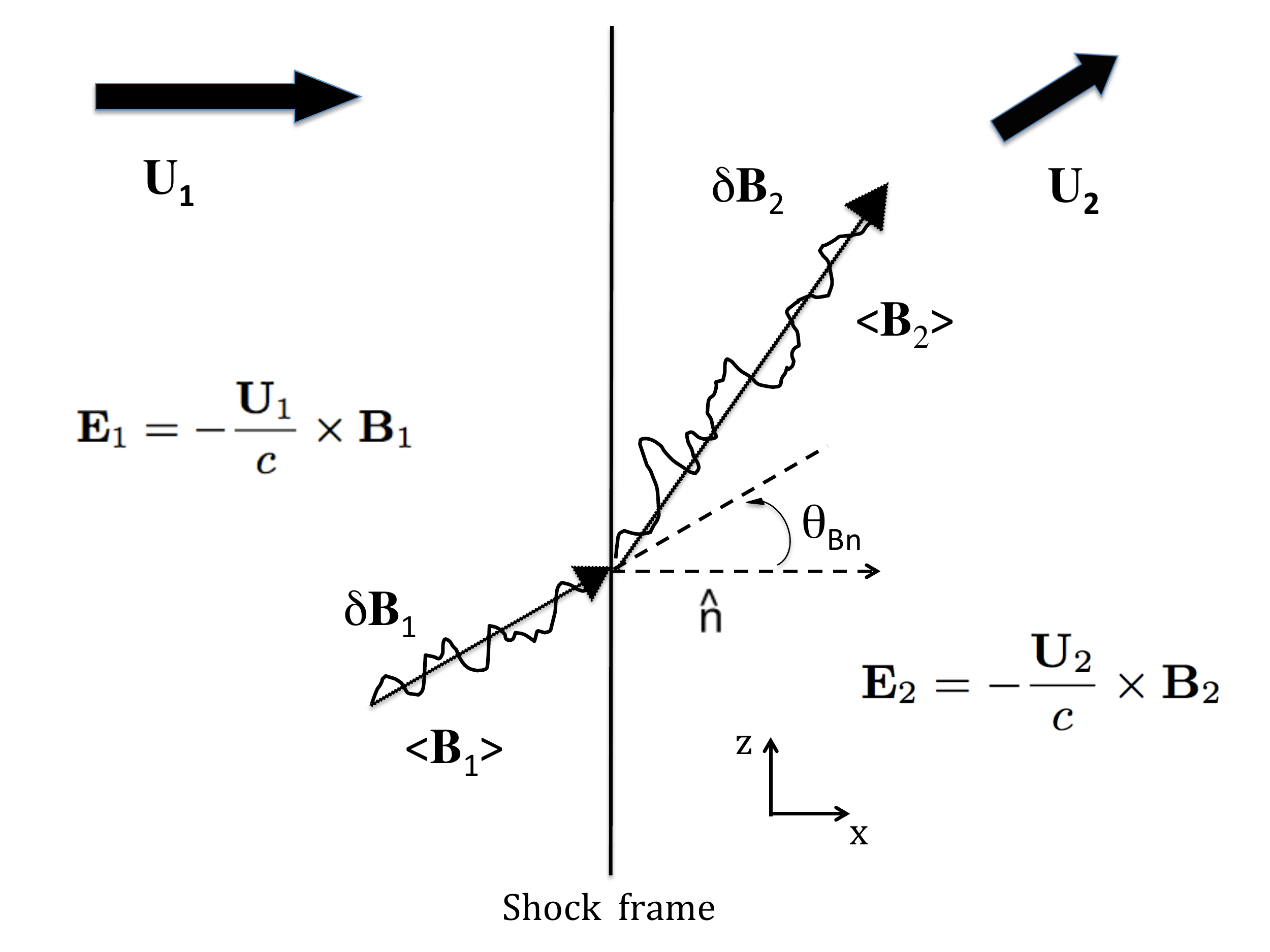}\\
\includegraphics[width=10cm]{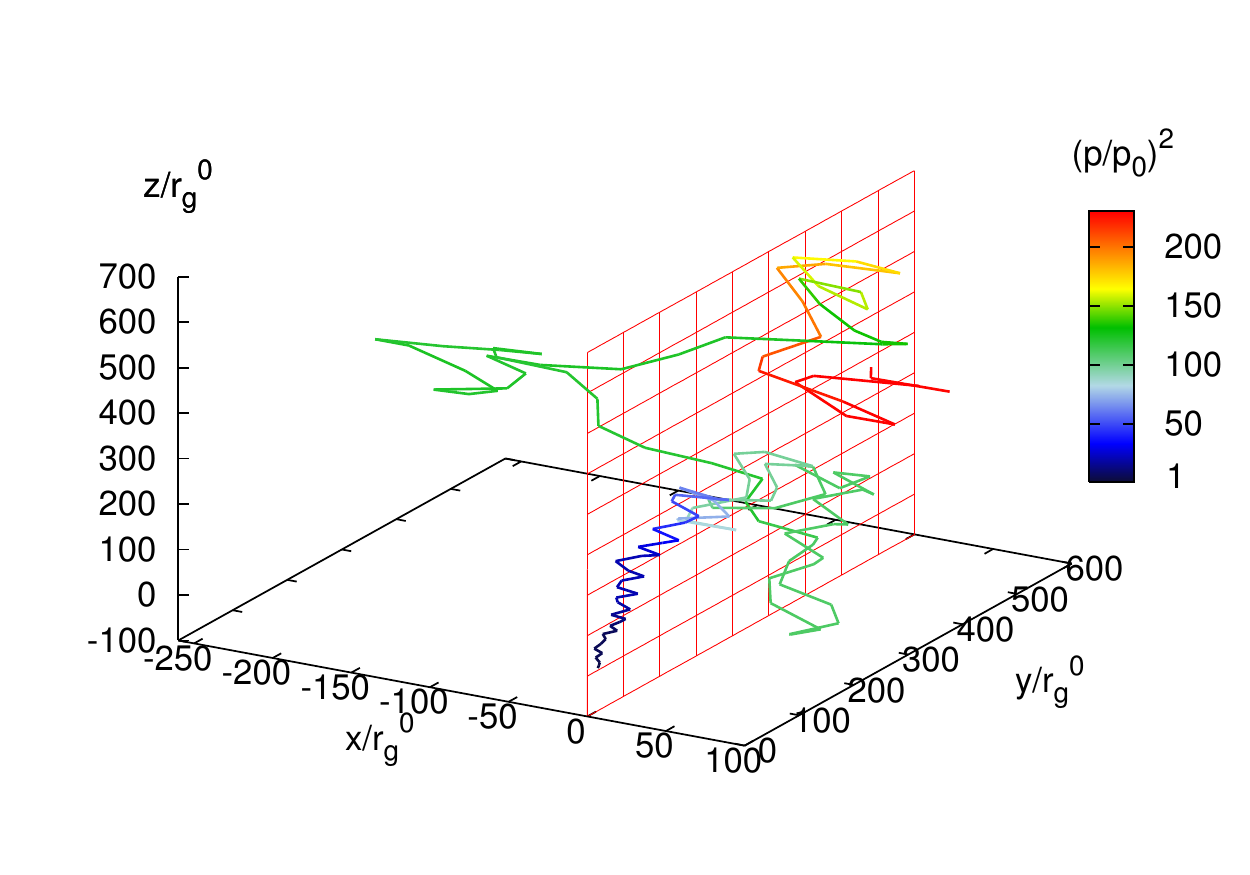}
\caption{{\it Upper panel} - Shock frame configuration of MHD variables (adapted from \citet{g05}). {\it Lower panel} - 3D trajectory of a sample particle injected at an oblique shock, propagating along $x$-axis, with $\theta_{Bn} = 85^\circ$ and $\sigma^2 = 0.3$. Spatial axis are in units of initial upstream gyroradius $r_g^0$. Color scale indicate increase of kinetic energy in the local plasma frame.}
\label{shock}
\end{figure}

The geometry used is depicted in Fig. \ref{shock}, upper panel.
The shock surface is at the plane $x=0$. Bulk plasma flows into the shock from $x<0$ to $x>0$ 
along the shock normal with speed $U_1 \ll c$, measured in the shock frame. Mass conservation 
across the shock implies a decrease of the bulk plasma speed from upstream to downstream along $x$ to $U_2 = U_1/r$,
where $r$ is the density compression at the shock. 
The profile of the bulk velocity is taken to be discontinuous at the gyroscale of the particles 
considered, much larger than the shock thickness, of order of ion skin depth; thus we use $U_x(x) = U_1$ for $x<0$, $U_x(x) = U_2$ for $x>0$ (see Appendix).

The upstream three-dimensional magnetic field is given by ${\bf B(x) = B}_0 + \delta {\bf B(x)}$, 
with an average component ${\bf B}_0 $ having orientation $\theta_{Bn}$ with respect to the shock normal 
and a random component  ${\bf \delta B} = {\bf \delta B} (x, y, z)$ having a zero mean ($\langle \delta {\bf B(x)} \rangle = 0$); the correlation length of the turbulence, $L_c$, is much greater than the particle gyroradius, $r_g$, at all energies considered in this paper. 

For the sake of simplicity, we assume that the three-dimensional power-spectrum of $\delta {\bf B}$ in the  inertial range, $G(k)$, is isotropic and scale-invariant, or Kolmogorov: 
$G(k) \propto k^{-\beta -2 }$, where $k$ is the wavenumber magnitude,
$\beta = 5/3$ is the one-dimensional power-law Kolmogorov index and the additional $2$ accounts 
for the dimensionality of the turbulence (see Appendix).

Non-thermal seed particles are injected at the shock with the same momentum magnitude, $p_0 = m \gamma_0 v_0$, and pitch-angle isotropic momentum distribution in the local plasma rest frame, where the cosinus of the pitch-angle with respect to the average field is defined as $\mu_{global} = {\bf p}\cdot{\bf B}_0/p B_0$. A sample particle trajectory is shown in Fig. \ref{shock}, lower panel. We solve numerically the Lorentz force equation for the charged particles in the prescribed magnetic field (see Appendix for details). For any $\theta_{Bn}$, the particle injection speed in the local plasma frame $v_0$ is comparable to the flow speed ($v_0 = 3.5 U_1$).  
The analytic solution of the induction equation is used in the particle equation of motion that  
is numerically solved with the procedure described in the Appendix. 
As usual, since the plasma is infinitely conductive, i.e., in the plasma rest frame 
both upstream and downstream the electric field vanish, 
particles undergo acceleration by the shock frame electric field ${\bf E}_\perp (x, y, z, t) = - {{\bf U}(x) / c} \times {\bf B}_\perp(x, y, z, t) $, where ${\bf U} (x) = (U_x, 0, U_z)$ and $\perp$ indicates the plane orthogonal to the shock frame flow velocity ${\bf U} (x)$; in Appendix ${\bf B}(x, y, z, t)$ is related to ${\bf B}(x_0, y, z, t_0) = {\bf B}_0 (x_0) + \delta {\bf B}(x_0, y, z, t_0)$. We neglect the electric field of the order of $v_A B/c$ arising from the magnetic field fluctuations modelled as Alfv\'en waves propagating along the field at speed $v_A$, as measured in the local plasma frame (see \citet{sv99} for a review on the limits of the magnetostatic approximation). This is justified by the fact that for the parameters used in our test-particle simulations the injection particle speed is much greater than Alfv\'en speed on both sides of the shock (see Sect. \ref{profiles} for details). In this work we focus on the particle acceleration resulting from the fluid compression at the shock. Stochastic acceleration, relevant as the particles move away from the shock, is expected to have a smaller effect on the spike formation because $1)$ the electric field $v_A B/c$ is disordered and $2)$ the acceleration is efficient only in resonant condition, i.e., $r_g$ comparable with length-scale of the turbulence, not reached in our simulations (the study of the stochastic acceleration is deferred to a separate work).

In order to obtain a steady-state solution, we use an approach similar to that in \citet{g05} (see Appendix).
We do not assume a free-escape boundary upstream but, instead, follow particles until either {\it a)} they reach a pre-specified high-energy cut-off or {\it b)} escape by advection downstream.  
Energetic particles, once in the upstream medium, undergo scattering on the pre-existing magnetic turbulence which advects with the upstream plasma. Our assumed magnetic power spectrum includes only the pre-existing fluctuations and not those that are self-generated by the particles. 

The sought spatial structure of particle density across the shock has a scale of hundreds of supra-thermal particle gyroradii. Thus, the test-particle approach can easily capture the proton scales we are interested in; also, the thickness of the shock dissipation layer (of order of ion skin depth) is negligible with respect to the gyroradius of the supra-thermal particles.

The relevant parameters are the ratio of the particle gyroradius 
to the turbulence correlations length, $r_g/L_c$, the normalized upstream turbulence energy density, 
$\sigma ^2 = (\delta B / B_0)^2$, the upstream magnetic obliquity $\theta_{Bn}$.  
Therefore our treatment applies to energetic particles at astrophysical shocks 
in various contexts: from the interplanetary, to the supernova remnant, to the intergalactic medium.

\section{Particle trajectories}\label{trajectories}

\begin{figure}
\includegraphics[width=8cm]{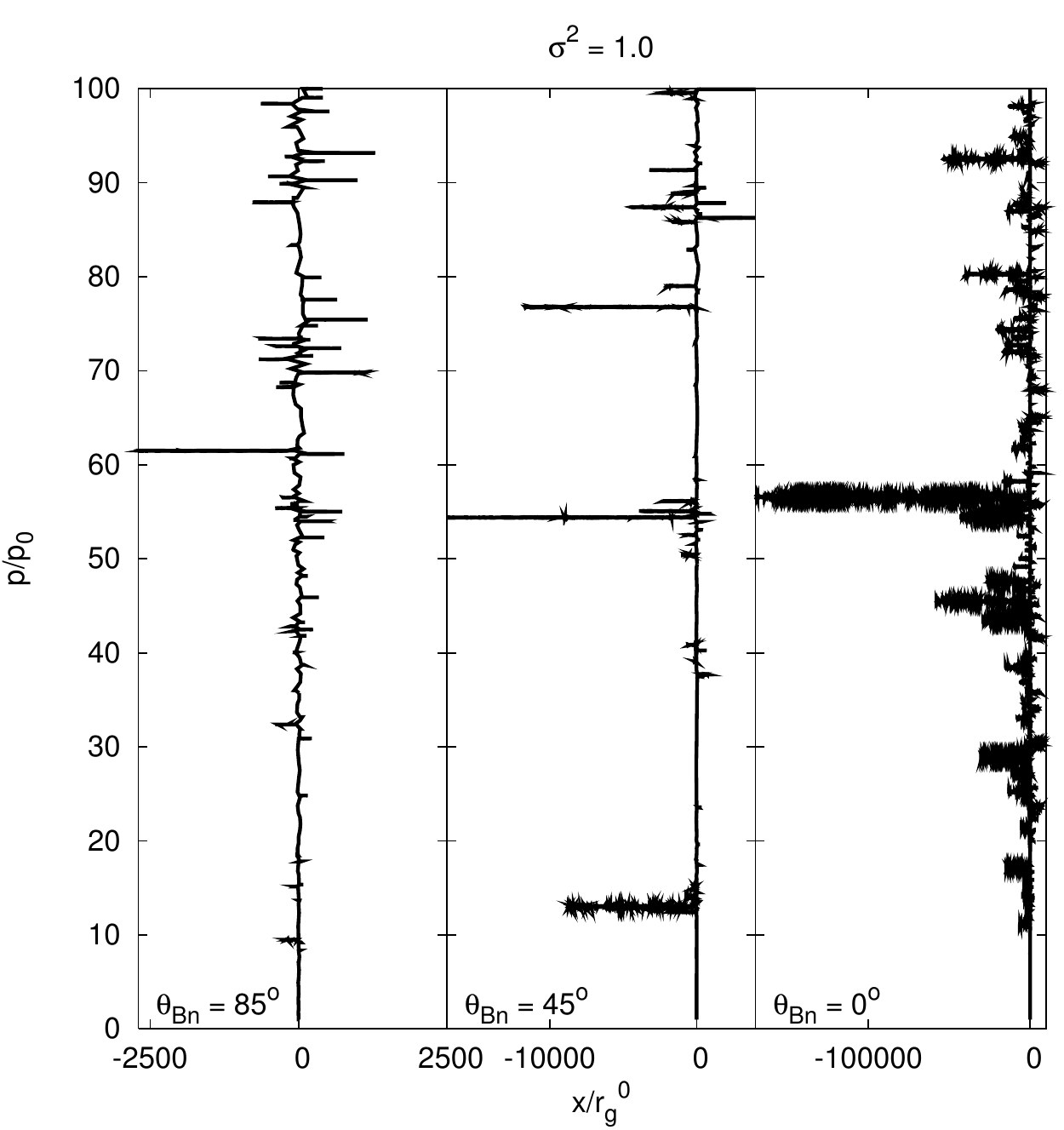}
\caption{Momentum in units of injection momentum $p_0$ versus the $x$-coordinate for three particles injected with random velocity at the shock, located at $x=0$, for various shock obliquities $\theta_{Bn}$ ($\sigma^2 = 1.0$). Length is in units of initial upstream particle gyroradius; notice the difference in spatial scale for various $\theta_{Bn}$. Upstream medium is to the left. Orbits show multiple shock crossing.}
\label{PX_1d0}
\end{figure}

In the present section, we show individual particle trajectories to illustrate salient features of the particle energization at shocks with various obliquity; in the following section we will describe the resulting spatial profile of a mono-energetically injected population of supra-thermal particles, with random pitch-angle and phase. 

We have numerically integrated the proton trajectories for various $\theta_{Bn}$ and $\sigma^2$.
Figure \ref{PX_1d0} traces the actual trajectories of sample protons injected at shock for three different $\theta_{Bn}$ and $\sigma^2 =1.0$. The injection particle speed in the local (downstream) plasma frame $v_0$ is only $3.5$ times the speed of the shock in the upstream plasma frame $U_1$ for every $\theta_{Bn}$. Particles are energised preferentially along the shock surface, regardless of the value of $\theta_{Bn}$, suggesting that drift acceleration plays a significant role even at shocks with $\theta_{Bn} = 0^\circ$. If the pre-existing upstream magnetic fluctuation is sufficiently strong ($\sigma^2 = 1.0$, Fig.\ref{PX_1d0}), making $\delta \vect{B}$ spatially isotropic, the turbulent motional electric field $-\vect{U}/c \times \delta \vect{B}$ has a component also along the shock surface and contributes to accelerate particles even at $\theta_{Bn} = 0^\circ$. If $\delta B$ is weaker, 
the particle spends more time far from the shock: the acceleration time is given by $ \tau_{acc} \sim \kappa_x/U_1^2$ \citep{drury83}, where $\kappa_x$ is the spatial diffusion coefficient along the direction normal to the  shock.  
From quasi-perpendicular to parallel shock (the latter requiring longer elapsed time than the former as $\kappa_{\perp} \ll \kappa_\parallel$), the particle excursion far upstream of the shock increases by two orders of magnitude due to the advection of the magnetic field lines dragging the particle back to the shock for $\theta_{Bn} > 45^\circ$. 

As for small momenta ($p \lesssim 10 p_0$) as seen in Fig. \ref{PX_1d0},  
a parallel shock confines more efficiently particles around the shock through an enhanced scattering if $\sigma^2=1.0$ than $\sigma^2=0.3$. 
In the next section we confirm that for $\sigma^2=1.0$ the spike at a $\theta_{Bn} > 0^\circ$ shock is higher (relative to the downstream background) and persists at high momentum compared to $\sigma^2=0.3$: high $\sigma^2$ increases the permanence time of particles at shock, thus increasing the contribution of the shock crossing to the particle acceleration. Conversely for high-obliquity shocks: the permanence time within the spike increases for smaller $\sigma^2$ as spike is generated by the SDA nearly scatter-free process (cfr. Fig.\ \ref{X_90_03_comp_p}), as first modelled in \citet{d83}:  particle slides along $y-z$ coordinates under the turbulent electric field $-\vect{U}/c \times \vect{B}$ ($\vect{B} = \vect{B}_0 + \delta {\bf B}$) while remaining close to the shock. 

\section{Supra-thermal particles density profiles}\label{profiles}

\begin{figure}
\includegraphics[width=8.5cm]{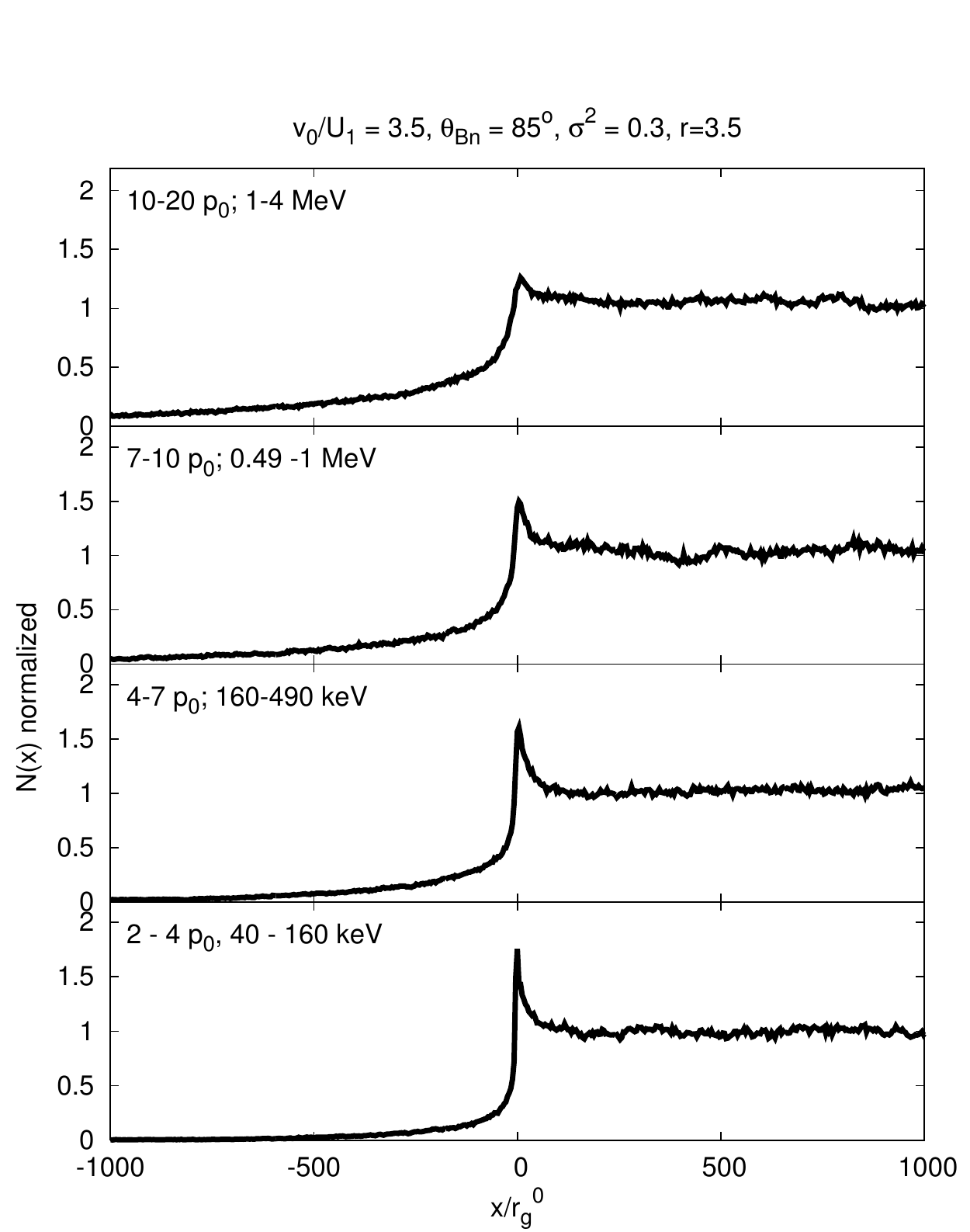}
\caption{Steady-state spatial density profiles of accelerated particles for various momentum intervals, and correspondent proton energy, at $\theta_{B_n} = 85^\circ$ and $\sigma^2 = 0.3$. The density is normalised to the far downstream value. The distance is in units of initial upstream gyroradius at $1$ AU.}  \label{X_90_03_comp_p}
\end{figure}

\begin{figure}
\includegraphics[width=8.5cm]{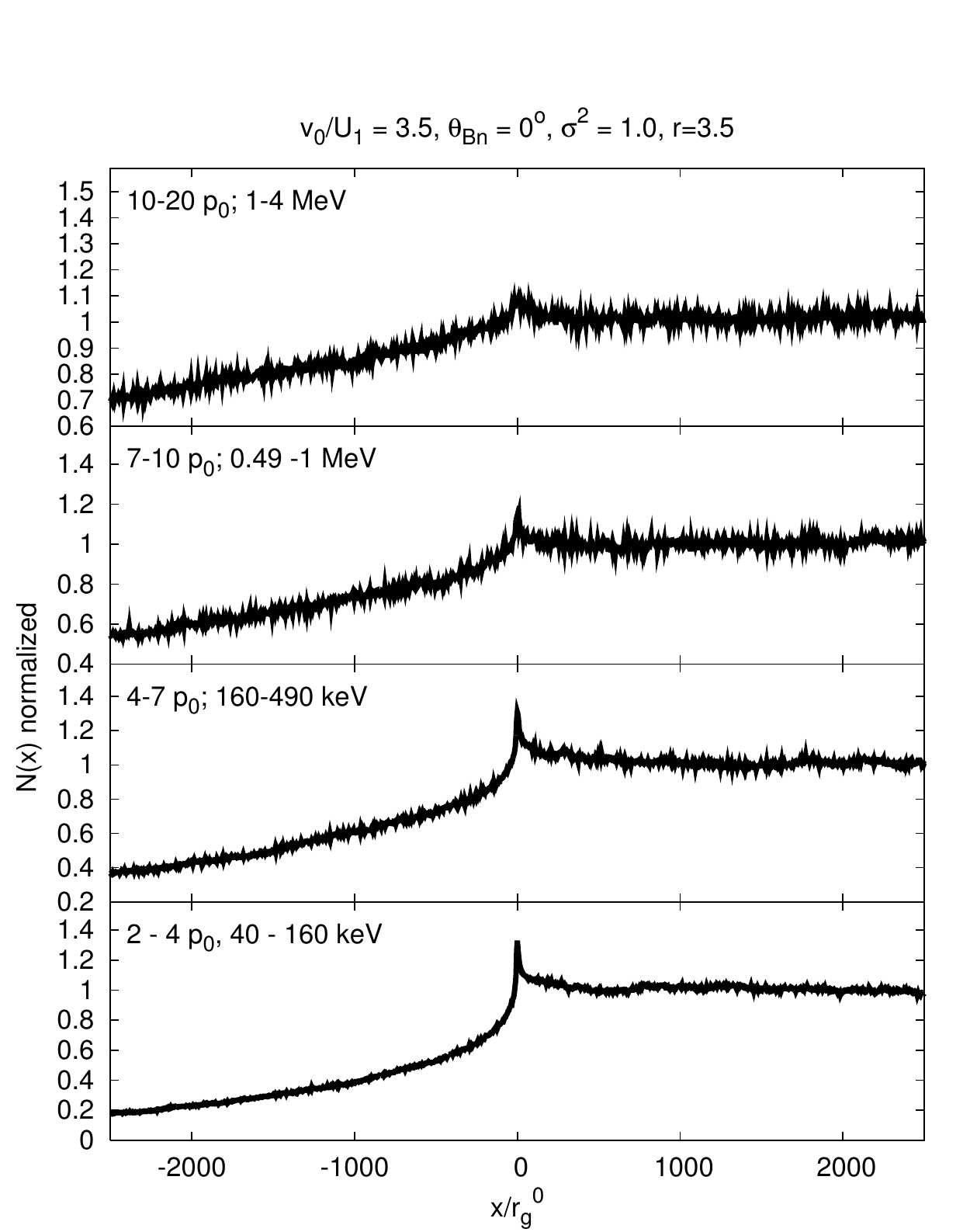}
\caption{Steady-state spatial profiles for various particle momentum intervals at $\theta_{B_n} = 0^\circ$ and $\sigma^2 = 1.0$ (see Fig.\ref{X_90_03_comp_p}). } \label{X_0_comp_p}
\end{figure}

\begin{figure}
\includegraphics[width=8.5cm]{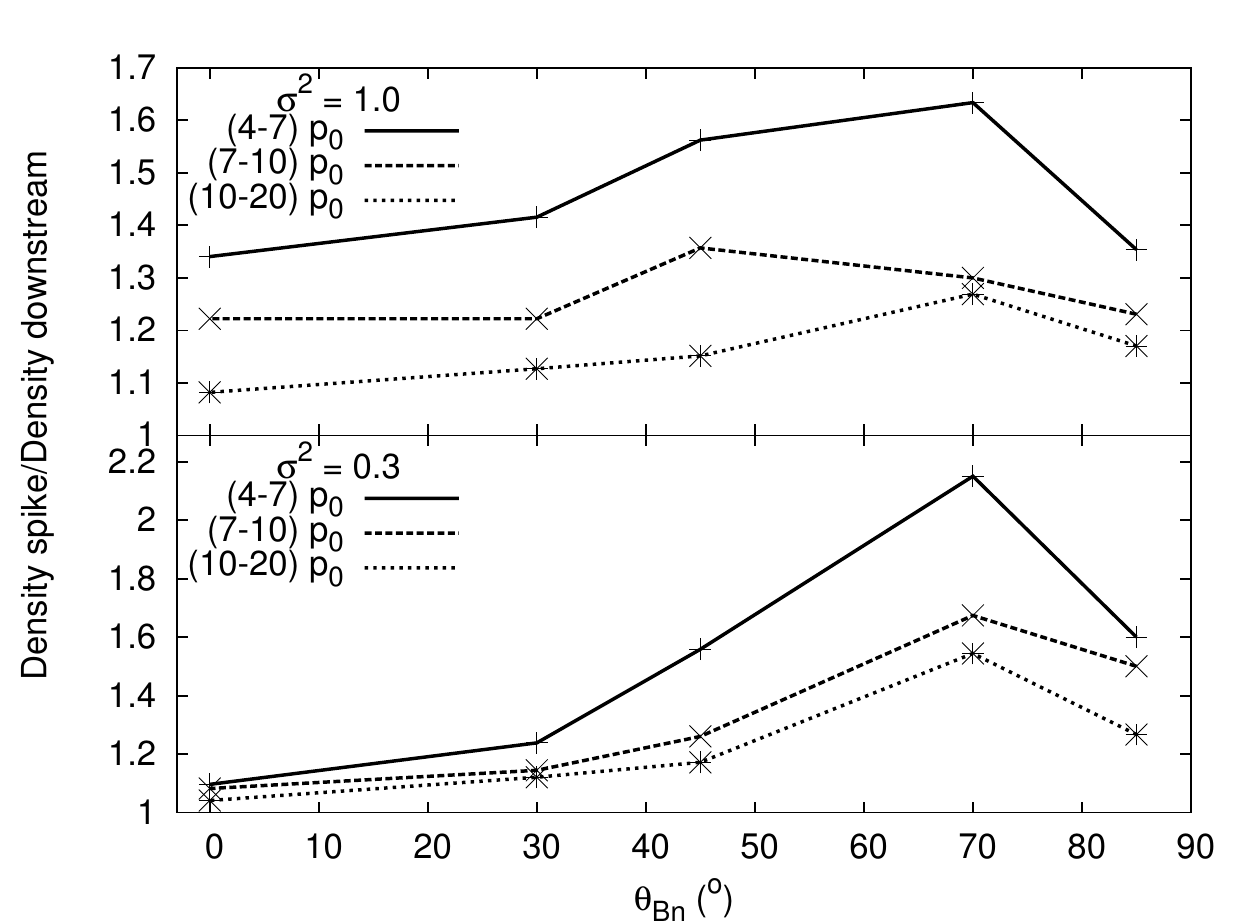}
\caption{Density enhancement at the shock in units of far downstream value as a function of $\theta_{Bn}$ for various $\sigma^2$ and plasma-frame particle momentum. Note the difference in scale of the vertical axis.  \label{spike_multi}}
\end{figure}

In this section we present the steady-state spatial density profiles $N(x)$ for accelerated protons from our numerical simulations. We show $N(x)$ at various momenta as measured in the local plasma frame to be compared directly to the solution of the steady state transport equation. Particles are injected with $v_0/U_1 = 3.5$. A strong shock is used (plasma density compression is  $r = 3.5$) with Alfv\'en Mach-number $M_A = 7$. The downstream boundary $x_b = 7.5 \times 10^3 r_g^0$ (see Appendix) is chosen to ensure that the highest-energy particles have enough space in the downstream medium to isotropize in pitch-angle. As a momentum boundary (described earlier), we use $p_b = 500 p_0$. For a strong shock travelling at $1$ AU in the solar wind, the ratio $v_0/U_1 = 3.5$ corresponds to a proton with kinetic energy $10$ keV at a shock propagating with $U_1 = 400$ km s$^{-1}$ in the upstream plasma  frame with an embedded  background magnetic field $\langle B_1 \rangle = 5$ nT ($r_g^0  = 2,900$ km and $\Omega_0 = e\langle B_1 \rangle/mc = 0.48$ s$^{-1}$); the upstream magnetic fluctuation has a correlation length $L_c = 0.01$ AU. Note that the momentum boundary $p_b$ is chosen so that highest-energy particle gyroradius in the upstream is smaller than the $L_c$. 

Figures \ref{X_90_03_comp_p}, \ref{X_0_comp_p} show numerically computed spatial density profiles of accelerated particles across the shock for various plasma-frame momenta. Unlike the prediction of the standard diffusion theory, we find a local enhancement (even more than two-fold) at the shock for low-energy accelerated particles at both quasi-perpendicular shocks ($\theta_{Bn} = 85^\circ$, Fig. \ref{X_90_03_comp_p}) and parallel shocks ($\theta_{Bn} = 0^\circ$, Fig. \ref{X_0_comp_p}). Note that the quasi-perpendicular shock profiles are shown for weak turbulence ($\sigma ^2 = 0.3$) and the parallel shock profiles are shown for strong turbulence ($\sigma ^2 = 1.0$), respectively in Fig.s \ref{X_90_03_comp_p}, \ref{X_0_comp_p}; those $\sigma^2$ values maximise the amplitude of the shock spike (see also Sect. \ref{anisotropy}). The amplitude of the spike decreases for increasing particle energy in both cases, as expected in the standard DSA theory for sufficiently fast particles: at higher momentum ($v \gg U_1$) the spike vanishes and the uniform downstream density is recovered. The length-scale of the intensity decay ahead of the shock depends on the spatial diffusion coefficient along $x$-axis (for the same $U_1$) thus extends less for the quasi-perpendicular than the parallel shock (typically $\kappa_\perp \ll \kappa_\parallel$). However, the presence of a shock spike might modify the exponential decay upstream: the comparison of the density profile upstream with the diffusion prediction and local non-diffusive behaviour is deferred to a separate work.

Our simulations contrast with results from other analyses (see, e.g., \citet{l07}) which suggest that only near-perpendicular shocks should exhibit a shock spike and the spike at parallel shock would be an artifact of the FTE method. In quasi-perpendicular case our downstream tail of supra-thermal particles density arises from the quasi scatter-free regime ($\sigma^2 =0.3$), as shown in the anisotropy analysis in Sect.\ref{anisotropy}, whereas in the parallel shock it results from isotropic scattering ($\sigma^2 =1.0$). An additional contrast is the qualitative structure of the spike: we find a trapping tail in the downstream density with thickness of order $ 10^2 r_g^0$, whereas FTE solutions or Monte-Carlo simulations sharply drop from the shock spike to the far downstream limit. The thickness of the downstream tail increases from perpendicular to parallel case, as a consequence of $\kappa_\perp \ll \kappa_\parallel$. We conclude that the deviation from DSA found here has different physical origin from the one discussed in previous works. Such a refined shock structure is observable by future interplanetary shock missions as discussed in Sect. \ref{helio}. 

Spike enhancements for various $\theta_{Bn}$, $\sigma^2$ and particle momenta are summarised in Fig. \ref{spike_multi}. At high-obliquity shocks ($\theta_{Bn} \geq 70^\circ$), the shock spike is much more enhanced at weak than strong turbulence, for every momentum, suggesting that the weaker is the fluctuating component of ${\bf B}$ the higher is the shock enhancement, that originates predominantly in a scatter-free drift acceleration limit, i.e., for uniform and static electric and magnetic field. Such an interpretation is corroborated by the pitch-angle anisotropy in the following section. The drop in enhancement between $\theta_{Bn} = 70^\circ$ and $85^\circ$ can be attributed to a more efficient advection of particles away from the shock as $\theta_{Bn}$ approaches $90^\circ$ than at $\theta_{Bn} = 70^\circ$.  Conversely, at parallel shock the spike amplitude is increased by strong turbulence.  Despite the different particle acceleration processes dominating at different $ \theta_{Bn}$, a downstream spike emerges at the same scale ($\sim 10^2 r_g^0$). For large $\theta_{Bn}$, the relative importance of the SDA regime decreases as $\sigma^2$ is increased. For small $\theta_{Bn}$ scattering on the turbulence on both sides of the shock is the main accelerating agent as maintains particles in the shock vicinity; thus the spike is enhanced at large $\sigma^2$. We therefore argue that the FTE solution should find a more prominent spike at stronger turbulence ($\sigma^2$ approaching $1$).

Our steady-state density profiles are not qualitatively affected by the shock strength. 
We performed simulations with density compression at the shock smaller than $3.5$ ($2.5$ and $1.5$) and found no qualitative difference in the spike enhancement; on the other hand, smaller compression leads to softer downstream momentum distribution (see Sect.\ref{spectra}), as expected from DSA, needing a larger particle pool.

\section{Anisotropy}\label{anisotropy}

\begin{figure*}
\includegraphics[width=17cm]{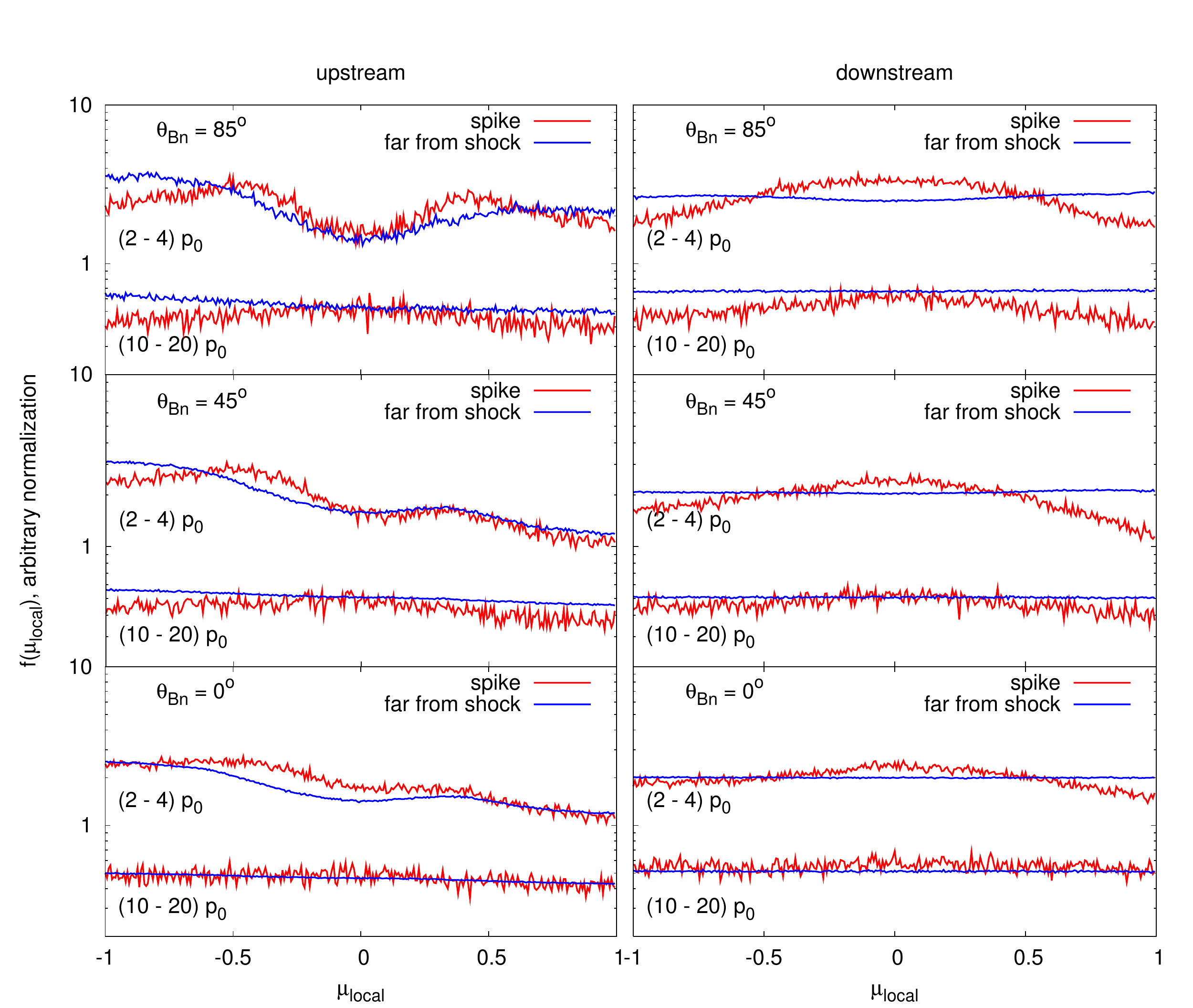}
\caption{Comparison of the steady-state pitch-angle $\mu_{local}$ distribution in the plasma frame in the far upstream (downstream) flow within two shells of thickness $30 r_g^0$ upstream (downstream) of the shock for charged particle propagating in a strong turbulence ($\sigma^2 = 1.0$) at various magnetic obliquity ($\theta_{Bn} = 85^\circ, 45^\circ, 0^\circ$ in various rows). Here $v_0/U_1 = 3.5$. The panels on the left (right) column refer to the upstream (downstream) fluid. In every panel the red (blue) curve corresponds to the fluid at the spike (far from the shock). Two ranges of momentum are considered ($[2-4], [10-20]$ in units of $p_0$) to show the progressive flattening of the excess at $\mu_{local} = -1$ (upstream) and of the ``pancake'' at $\mu_{local} = 0$ (downstream) for increasing energy. At high momentum, $p > 20 p_0$, the pancake disappears and $f(\mu_{local})$ collapses to an isotropic distribution, as expected from DSA. The relatively large numerical fluctuations at $\theta_{Bn} = 85^\circ$ upstream (top left panel) originate from the upstream lower statistics for a quasi-perpendicular shock. \label{MU_all_1}}
\end{figure*}

\begin{figure}
\includegraphics[width=9cm,height=12cm]{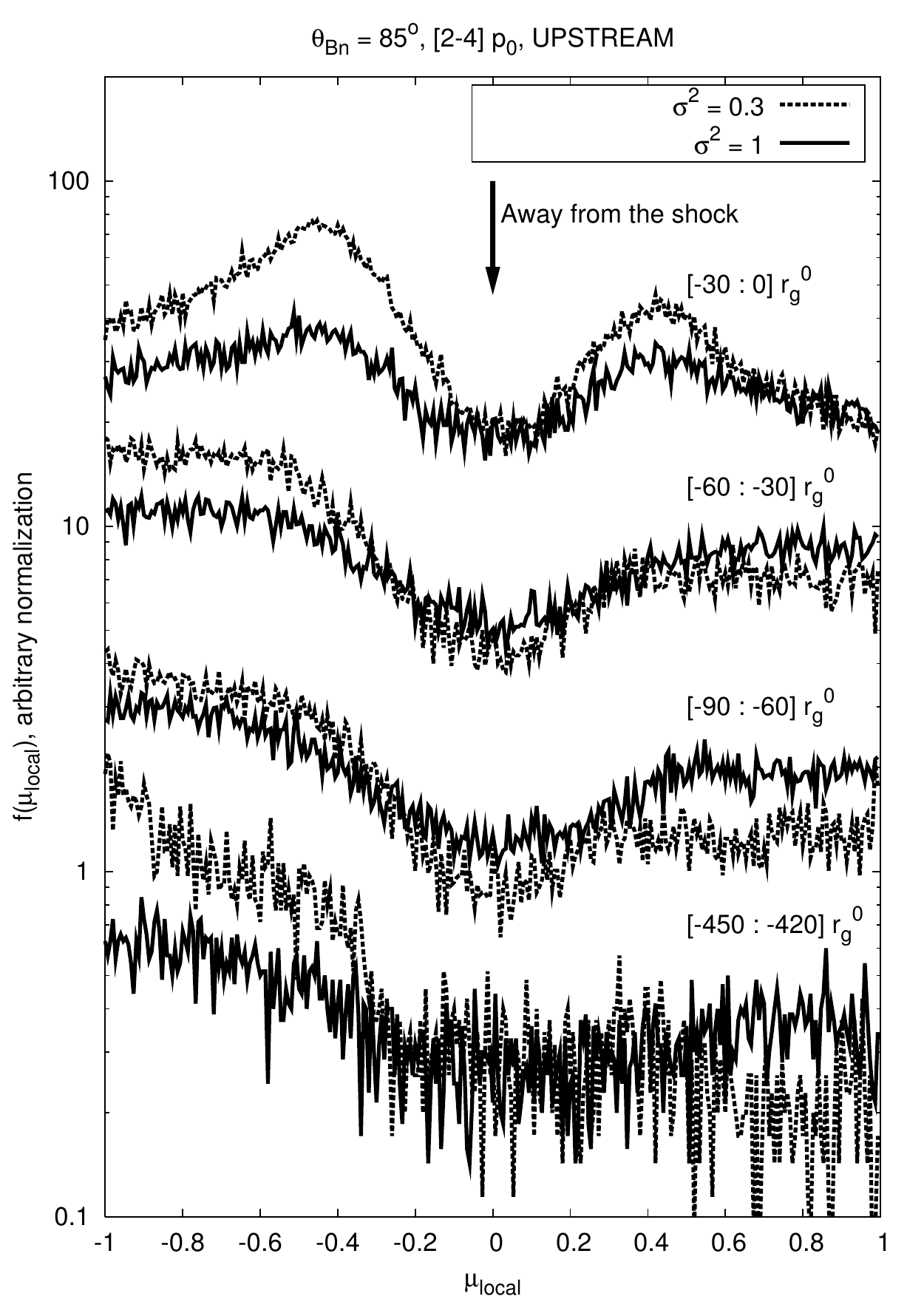}
\caption{Synopsis of the steady-state pitch-angle $\mu_{local}$ distribution in the upstream  plasma at various shells of $30 r_g^0$ thickness at increasing distance from the shock, downward from $[-30 : 0] r_g^0$ as far as $[-450 : -420] r_g^0$. Two different turbulence strengths are compared ($\sigma^2 = 0.3$ dotted line, $\sigma^2 = 1$ solid line). Here $\theta_{Bn} = 85^\circ$, $v_0/U_1 = 3.5$ and local plasma frame momentum in the range $[2-4]p_0$.}
\label{MU_varB}
\end{figure}

In this Section we compare at various locations the plasma-frame pitch-angle anisotropy by using a pitch-angle defined with respect to the local magnetic field at the particle position, with no spatial average: $\mu_{local} = {\bf p} \cdot {\bf B}/pB$. The distribution $f(\mu_{local})$ is depicted in four spatial regions in each row of Fig.\ref{MU_all_1}: two shells of thickness $30 r_g^0$ on both sides of the shock, far upstream and downstream\footnote{For a proton injected at kinetic energy $E = 10$ keV in the solar wind at $1$ AU, the shell has thickness $30 r_g^0 \sim 6 \times 10^{-4} $ AU.}. In Fig.\ref{MU_all_1} we consider two distinct momentum intervals ($[2-4], [10-20] p_0$), three magnetic obliquities ($\theta_{Bn} = 85^\circ , 45^\circ, 0^\circ $) and take $\sigma ^2 = 1.0$. The far downstream distribution is calculated in the interval between $0.1 x_b$  and $x_b$, which is a region much larger than the particle mean free paths so that the distribution is isotropic. The far upstream distribution is calculated in $x < -30 r_g^0$ to seek excess of reflected particles, whose pile-up was argued by \citet{gkga99} to originate the spike, that they found located solely in the upstream region. Such a thickness is broad enough to account for various particle populations: low-energy particles drifting along the shock in a multiple encounter\footnote{It is customary to define single encounter with a shock the period of time during which the particle remains within one gyroradius from the shock.} interaction, accelerated particles advected downstream, particles coming back to the shock from far downstream, particles reflected one or multiple times by the shock moving to the upstream and particles advected back to the shock from upstream. 

We find that in the downstream shell (Figs. \ref{MU_all_1}, right column, all panels, red curves) the lower momentum ($[2-4]p_0$) pitch-angle distribution has a ``pancake'' shape in logarithmic scale (broad bump peaking at $\mu_{local} = 0 $, i.e., for particles moving in the direction perpendicular to the local magnetic field), whereas in the far downstream fluid $f(\mu_{local})$ is closer to isotropic, as predicted by DSA. Likewise, within the scatter-free limit ($\sigma^2 = 0$) at quasi-perpendicular shocks a distribution peaked at $\mu_{global} = 0$ emerges (see the $\mu$-histograms after a single shock encounter in Figs. 7-17 of \citet{d88}). Such a peak is broadened by small-amplitude perturbations $\sigma^2 = 0.09$ \citep{d88} approaching the pancake-shape we see in our results. Far downstream (Fig.\ref{MU_all_1}, right column, blue curves) the distribution is featureless: the scattering is frequent enough that the distribution is isotropic. We emphasise that comparable pancake anisotropy in the downstream region is found for each $\theta_{Bn}$ (Fig.\ref{MU_all_1}) in the same momentum ranges in which a spike is seen; this suggests that such a local feature depends on local relative orientation of field line and shock surface and not by the value of the global obliquity $\theta_{Bn}$.

Figure \ref{MU_all_1}, left column, shows in detail the upstream anisotropy associated with the spike: particles spending a few gyroperiods in the thin shell ahead of the shock before being advected back to the shock or particles diffusing within the upstream shell and being accelerated. Upstream at the spike ($-30 r_g^0 < x <0$, Fig. \ref{MU_all_1}, left column, red curves), we find two bumps at $\mu^*_{local} \simeq \pm 0.45$ for each $\theta_{Bn}$. In order to interpret the bumps, we compare Fig. \ref{MU_varB} with the adiabatic test-particle theory and previous simulations in weak turbulent field \citep{d88}. We should point out that this is expected to be a qualitative argument as the adiabatic test-particle theory is based on the conservation of the particle's magnetic moment before and after a single encounter with a quasi-perpendicular shock whereas our simulations include both single and multiple encounters and track the pitch-angle also {\it during} the shock encounter. The lowest curve in Fig. \ref{MU_varB} shows at a distance between $[-450] r_g^0$ and $[-420] r_g^0$ an excess at $\mu_{local} = -1$ with an increasing anisotropy for smaller $\sigma^2$; similar broadening was found in the comparison of the scatter-free limit with the case $\sigma^2 = 0.09$ (see Fig. 17 in \citet{d88})\footnote{For the parameters chosen here, at all energies, the adiabatic test-particle theory provides a strong anisotropy for upstream reflected particles receding from the shock ($f(\mu)$ non-vanishing for $\mu < - 0.81$)}. However, by sampling the distribution of $\mu_{local}$ closer to the shock (Fig. \ref{MU_varB}), we find a stronger departure from the scatter-free limit. The bumps result from the turbulence that deforms the field lines locally at the shock front. Fluctuations allow a single field line to have multiple connection points along the shock front, leading to a bidirectional streaming and bumps in both hemispheres as reflected particles can come back to the shock. The bump at $\mu_{local} < 0$ is higher than $\mu_{local} > 0$ as a result of the excess of particles moving anti-parallel to the shock upstream in the scatter-free limit.
In the upper curves (at $[-30:0] r_g^0$) the bumps are enhanced and narrower for small $\sigma^2$, although we point out that  the $\mu_{local}^*$ is greater than the limiting anisotropy in the scatter-free limit ($\mu < -0.81$). Thus, Fig. \ref{MU_all_1} shows that the presence of two peaks and $\mu_{local}^*$ is independent on $\theta_{Bn}$. This corroborates our interpretation that the spike originates from the interaction of the small-scale magnetic field with the shock: in the presence of fluctuations the large-scale, or average, magnetic obliquity $\theta_{Bn}$ is not as important  close to the shock and the particle anisotropy is described by $\mu_{local}$ only.

At higher momentum ($[10-20]p_0$), the thickness of the shell is too narrow to allow for pitch-angle isotropization: for instance, the red curves in Fig.\ref{MU_all_1} are calculated in a shell of thickness $30 r_g^0$, i.e., only $1.5$ times broader than the initial gyroradius for particles with momentum $20 p_0$, insufficient to provide particle scattering for isotropization even for a mean free path as small as the gyroradius, i.e., Bohm regime.

The use of a pitch-angle with respect to the local field direction requires an additional justification.
If we consider the parallel shock case (Fig.~\ref{X_0_comp_p}),
the mean free path $\lambda_\parallel \simeq \kappa_\parallel / v$, 
with $v$ particle speed in the local plasma frame and $\kappa_\parallel$ estimated from
quasi-linear theory\footnote{Same conclusion is found if $\kappa_\parallel$ is numerically estimated from the exponential roll-over of the upstream particle density.}, is larger or comparable to spike thickness,
so within the spike the particles cannot isotropize in the pitch-angle, calculated with respect to ${\bf B}_0$, i.e. 
$\mu_{global} = {\bf p} \cdot {\bf B}_0/pB_0$. In other terms the steady state $f(\mu_{global})$ calculated close to the shock might reflect the isotropy in $\mu_{global}$ used to inject particles at the shock rather than being a genuine isotropy produced by downstream scattering: particles are not expected to behave diffusively near the shock, within the spike. In addition, we find that behind the shock $f(\mu_{global})$ is flat (not shown in this paper), for $\theta_{Bn} = 0^\circ$ at the same low momentum ($[2-4] p_0$) which exhibits the spike. 
However, in our simulations we compute the effect of a fluctuating magnetic field on a collection of particles 
by summing up the instantaneous (on time-scale smaller than $\Omega_0 ^{-1}$) 
position and pitch-angle of the actual trajectories. This suggests that whether or not the distribution isotropizes through scattering must be evaluated at a smaller scale with $\mu_{local} = {\bf p} \cdot {\bf B}/pB$. Since we ensemble-average over several turbulence realizations and spatially average over uncorrelated regions of the shock surface, any systematic effect of the turbulence on the anisotropy should average out. 

\section{Spectra}\label{spectra}

\begin{figure}
\includegraphics[width=8cm]{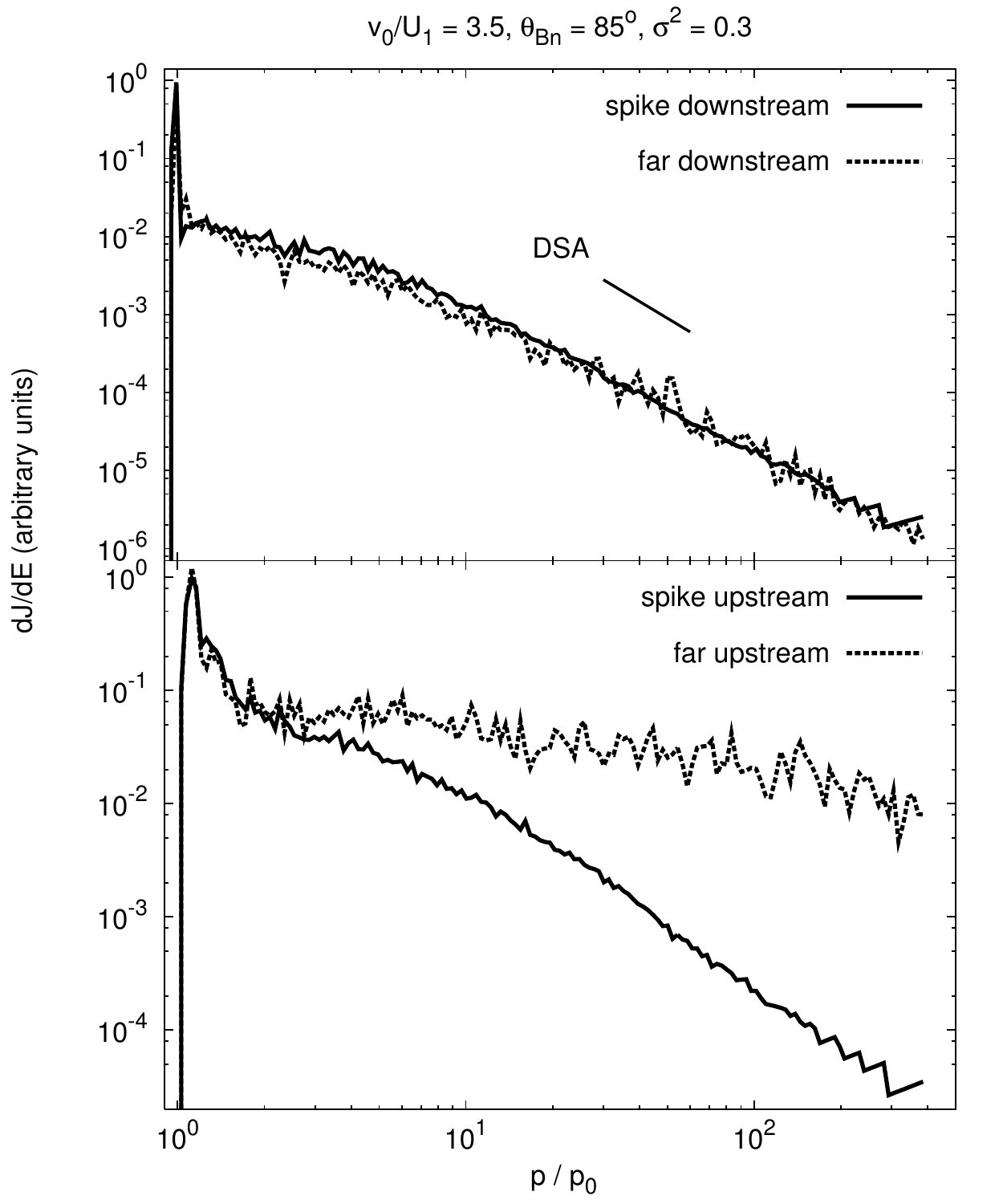}
\caption{Steady-state differential intensity spectrum as a function of momentum in the local plasma frame in various spatial regions (within the two shells on both sides of the shock and far from it) for $\theta_{B_n} = 85^\circ$, $\sigma^2 = 0.3$ and $v_0/U_1 = 3.5$ compared with DSA prediction far downstream. The top panel shows $dJ/dE$ downstream (within $0 < x < 30 r_g^0$ the solid line and $[0.1 : 1] x_b$ the dashed line). The bottom panel bear on the upstream flow (within $ - 30 r_g^0 < x < 0$ the solid line and $x < - 30 r_g^0$ the dashed line).
\label{P_85_03_comp_p}}
\end{figure}

Figure \ref{P_85_03_comp_p} shows the steady-state differential intensity spectra $dJ/dE = p^2 f(p)$ as a function of the momentum in the local plasma frame within various spatial regions compared with the downstream DSA prediction, i.e., test-particle limit power-law $p^{-\alpha+2}$, where $\alpha = 3r /(r-1)$ ($\alpha =4.2$ for $r = 3.5$). In the far downstream region the spectrum steepens to the DSA limit at smaller momentum for parallel than for perpendicular case (see also \citet{g05}): since typically $\kappa_\perp \ll \kappa_\parallel$ fewer energetic particles return to the shock, if highly oblique, to be further accelerated. A comparison of the spectra near the spike, but downstream of it, to that far downstream of the shock is shown in the top panel of Fig. \ref{P_85_03_comp_p} within the range $[3 -20] p_0$. There is a prominent $\mu_{local}$-anisotropy in this case (see Fig.\ref{MU_all_1}), leading to excess of low-energy particles at the spike with respect to the far downstream region\footnote{Such an excess can be quantified with a non-linear least-squares fit: for the case $\theta_{Bn} = 85^\circ$, $\sigma^2 = 0.3$ (Fig.\ref{P_85_03_comp_p}) we find slopes $-1.393 \pm 0.045$ (spike downstream) and $ -1.294 \pm 0.044$ (far downstream), compared with the DSA value $-2.2$ expected at higher momenta. For the case $\theta_{Bn} = 0^\circ$, $\sigma^2 = 1.0$, we find slopes $-1.680 \pm 0.022$ (spike downstream) and $-1.573 \pm  0.055$ (far downstream); clearly, the parallel shock is closer to DSA limit.}.
In both cases considered, the spectrum at the spike softens because the population piling-up at the spike is deplenished in the most energetic particles.

\section{Application to astrophysical shocks}\label{helio}

In this Section, we discuss the implications of our findings for the {\it in-situ} measurements of interplanetary shocks\footnote{We have performed simulations of supra-thermal protons at shocks of  galactic supernova remnant as well. We find a shock spike for injection energy $E = 10$ MeV at a strong shock ($r = 4$) propagating into the ISM ($B_0 = 3 \mu $G with small fluctuation, i.e., $\sigma^2 = 0.3$, $L_c = 2.2 \times 10^{15}$ cm) with $\theta_{Bn} = 85^\circ$ and $U_1 = 6,600$ km$/$s. However, the thickness of the spike is very narrow ($10^{-6}$ pc) compared with the typical radius of a few parsec for a young or middle-aged remnant.}.

\subsection{Low-corona shock}\label{lowcorona_shock}

\begin{figure}
\includegraphics[width=8.5cm]{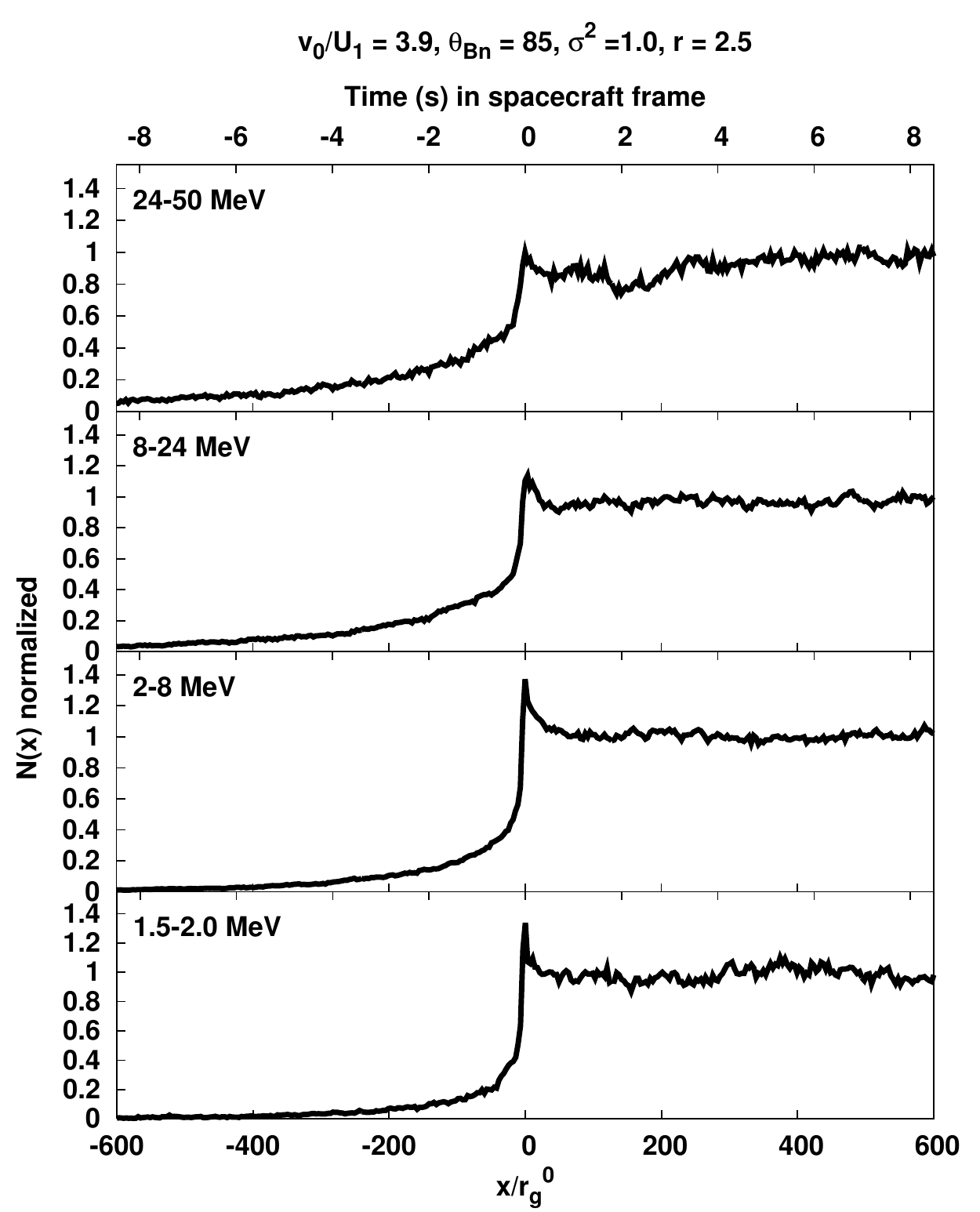}
\caption{Steady-state spatial density profiles of accelerated particles for various particle momentum ranges at $\theta_{B_n} = 85^\circ$ and $\sigma^2 = 1.0$. The density is normalised to the far downstream limit. In the lower $x$-axis the distance is in units of initial upstream gyroradius, in the upper the time is in units of a spacecraft past by the shock. The spike duration between $1.5$ and $8$ MeV amounts to ($\sim 1$ sec) ensuring observability by an energetic particle detector with current and future time-resolution. \label{LC_comp_p}}
\end{figure}

The detection of spikes at interplanetary shocks has been rare so far, due to limited time-resolution of the energetic particles detectors. We predict in Fig. \ref{LC_comp_p} the energetic particles profiles at travelling shocks moving at high speed close to the Sun. For protons injected at kinetic energy $E = 50$ keV in the slow solar wind at $10$ solar radii ($B_0 = 0.025$ G) at a relatively strong ($r = 2.5$) quasi-perpendicular ($\theta_{B_n} = 85^\circ$) shock we use $U_1 = 2,500$ km s$^{-1}$ ($M_A = 2$) and $L_c = 10^{-4}$ AU. The predicted duration of the shock spike as seen by a spacecraft crossing the shock along the $x$-axis turns out to be a few seconds (shorter for greater $\theta_{Bn}$ as shown previously). Finally, we find comparable amplitude and duration of the spike if the momentum is calculated in the local plasma frame or in the shock frame. {Instruments on the upcoming NASA mission Solar Probe Plus, including those that measure energetic particles \citep{mccomas14}, will have a cadence less than this and will be capable
of resolving the shock spike events.}

\subsection{Solar Termination Shock}\label{solar_term_shock}

\begin{figure}
\includegraphics[width=8.5cm]{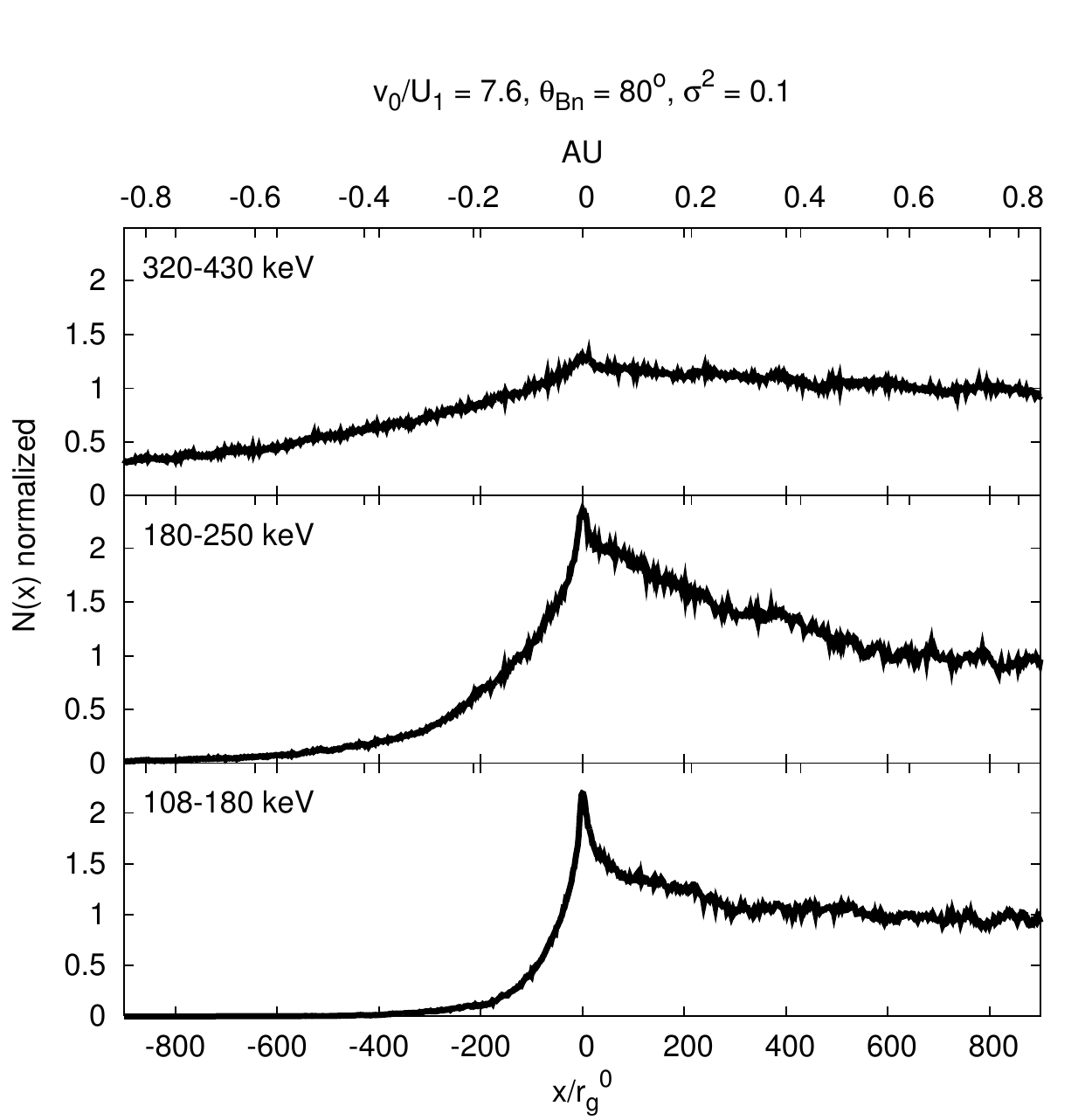}
\caption{Spatial density profiles of solar termination shock accelerated particles for various particle kinetic energy ranges at $\theta_{B_n} = 80^\circ$ and $\sigma^2 = 0.1$. The density is normalised to the far downstream limit. The distance is in the lower $x$-axis in units of initial upstream gyroradius, in the upper in units of AU. \label{TS2_100_comp}}
\end{figure}

The crossing of the solar termination shock (TS) by {\it Voyager-1} \citep{d05} reveals that energetic particles exhibited a spike in intensity at the time of shock passage in high-energy protons ($3.4 -17.6$ MeV) and ions ($40 -53$ keV). In Fig. \ref{TS2_100_comp} we display the simulated density profile with parameters compatible to the numerical solution of the FTE for $\theta_{Bn} = 80^\circ$ in \citet{fzl08}: $U_1 = 350 $ km s$^{-1}$, $r =3.0$, $v_0/U_1 {\rm cos} \theta_{Bn} = \sqrt{5/3}$; also, here $B_0 = 5 \times 10^{-7}$ G, $M_A = 10$ and $L_c = 0.01$ AU. Ions in $108-180$ keV (lower panel in Fig. \ref{TS2_100_comp}) exhibit a shock spike less prominent than the spike at $150$ keV found in \citet{fzl08}, i.e., a factor $10$ {\bf (a smaller spike is found  for smaller $\theta_{Bn} $)}. The finite energy interval chosen here, in contrast with a mono-energetic population in \citet{fzl08}, might increase the far downstream limit relatively to the spike, reducing the amplitude of the spike. However, we have used a 3D-isotropic magnetic turbulence power spectrum, in contrast with the slab turbulence used in \citet{fzl08}: as a general consequence of particle transport \citep{jkg93}, in a slab turbulence particles are unphysically forced to adhere to the field lines, i.e., to a direction lying on the shock surface (in the case of $\theta_{Bn} = 80^\circ$ considered in \citet{fzl08}), on a time-scale shorter than the scattering time \citep{fj11}. Thus, one would expect a less prominent spike in the slab case as the advection carries particles away from the shock. Although a slab might correctly approximate the Alfv\'enic turbulence at the TS, the isotropic turbulence used here allows to account for both perpendicular and cross field diffusion with no assumptions.

\section{Discussion and Conclusion}\label{discussion}

We have performed several test-particle numerical simulations by integrating the trajectories of individual energetic particles scattering off a magnetic turbulence advected through a planar shock to study the formation of spike-like structures, or localised intensity enhancements, not predicted by the standard steady-state DSA. Previous works based on numerical integration of the focused transport equation \citep{l07} argued that no local enhancements can be found at parallel, or small obliquity, shocks for low-energy suprathermal particles. Our direct particle trajectory integration shows that such spikes do indeed exist at quasi-parallel shocks that move through large-scale magnetic turbulence.

The presumed continuous transition from SDA to DSA as dominant scenario of particle acceleration at shocks as going from quasi-perpendicular to quasi-parallel geometry, although roughly consistent with several observations and models, contains discrepancies and remains to be understood (see \citet{l12} for a review in the context of interplanetary shocks). Our results aim at reconciling the dichotomy between the modelling of spike at quasi-perpendicular shocks with SDA \citep{d83}, and the argument for less oblique shocks ($\theta < 70^\circ$) that the shock enhancement arises from particles trapped in the upstream by self-generated hydromagnetic waves, with a resulting drop of particles mean free path \citep{s85}.

We find a local pitch-angle anisotropy resulting in a pancake-like shape, independent on the value of the large-scale obliquity $\theta_{Bn}$. Therefore it seems not necessary to assume the presence of self-generated waves as the turbulence, inherent property of the   medium travelled by shocks, is able to trap particles also at $\theta_{Bn} \sim 0^\circ$ shocks, at least at supra-thermal energies.

The approach of simulating the direct particle trajectory in magnetic turbulence used here naturally includes parallel and perpendicular diffusion to the large scale uniform magnetic field, without assuming a particular form of perpendicular transport.
We note that a decorrelation of low-energy particles might occur within a few gyroperiods time-scale even in weak turbulence ($(\delta B/B_0)^2 = 0.1$), as direct simulations in \citet{fg12} suggest. Suppressing the perpendicular diffusion, namely artificially constraining the particle motion (see for instance \citet{fj11}), might unphysically limit the spike amplitude in the quasi-perpendicular case because the advection along the flow artificially dominates over the perpendicular diffusion \citep{gj99}.

The spatial features at proton-scale sought here do not need hybrid simulations 
as test-particle approach can consistently capture the phenomenon we are interested in. 
However, in test-particle simulations the definition of magnetic obliquity has only large-scale validity,
because even at proton gyroscales the real shock structure will be corrugated by plasma micro-instabilities, 
so that the same particle reflected to the upstream 
will experience a different $\theta_{Bn}$ at its successive encounters with the shock.
The shock planarity is also affected by the density of the medium it is travelling into: 
large-scale corrugation induced by the upstream inhomogenieties will also
convert bulk kinetic energy of the shock to magnetic fluctuation power at the energetic proton gyroscale, reducing energetic particles mean free path, as demonstrated for high-Mach number non-relativistic shocks \citep{gj07,f13} and for mildly relativistic shocks \citep{m14}. Cross-shock electrostatic potential, generated by the charge separation between ions and electrons at the shock due to differential deflection in the magnetic field, has been neglected in this paper as confined to the electron skin-depth scale. Thus, the consistent study of electron density spatial profile, 
deferred to a future work, requires the use of hybrid simulations (particle electrons and fluid ions).

We have neglected the effect on the spike of the upstream self-generated waves. The amplitude of the self-generated turbulence
grows from perpendicular to parallel shock \citep{l12}. Therefore, should spike-events downstream 
in parallel shocks originate upstream from hydromagnetic waves trapping 
and accelerating particles which are further convected downstream and produce spikes,
one should find the largest amplitude spikes at parallel shocks.
However, in our simulations spikes at globally oblique shocks ($\theta_{Bn} \simeq 70^\circ$) reach the highest amplitude. This feature, not explained in our test-particle approximation, requires further investigation. 

In summary, since spikes are shown here to be an inherent property of shocks, 
we underline that, beside the {\it Voyager}-1 crossing of termination shock, 
a much larger sample is expected to be seen at interplanetary shocks by high-time resolution particle detectors. To this purpose, investigation of time-dependent spatial profiles, not considered in this work, might also help interpretation of data.
Our simulations might be relevant to the future Solar Probe Plus and Solar Orbiter Missions. Unprecedently high time-resolution in situ measurements in the solar atmosphere (high solar corona down to $\sim 10$ solar radii) will enable measurements of small-scale spatial features in energetic electrons and protons distributions 
at interplanetary shocks. This will contribute to clarify the behaviour of non-thermal particles at non-relativistic shocks,
one of the main focuses of Solar Probe Mission, possibly impacting on the understanding of 
energetic particles at supernova remnant shocks.

\section*{Acknowledgments}

We acknowledge the useful discussions with J. R. Jokipii and J. K\'ota. 
We thank the anonymous referees for useful suggestions and comments.
This work was supported, in part, by NASA under 
Grants NNX10AF24G and NNX11AO64G. Work by JG was also supported, in part, from the ISIS instrument on
NASA's Solar Probe Plus mission, and by the NSF under grants
AGS1135432 and AGS1154223. This work benefited from technical support 
by the computer cluster team at the Department of Planetary Sciences 
at University of Arizona.

\appendix

\section{Particle trajectory integration}

The equation of motion of a test-particle with charge $e$ and mass
$m$ moving with velocity $v (t)= |\vect{v} (t)|$ in a magnetic field $\vect{B}(\vect{x})$ is the Lorentz equation, written in the shock frame as
\begin{equation}
\frac{d\vect{u}(t)}{dt} =  {\cal E}(\vect{x},t) +  \frac{\vect{u}(t) \times\vect{\Omega}(\vect{x},t)}{\gamma(t)}  \;,
\label{lorentz}
\end{equation}
where we defined $\vect{\Omega}(\vect{x},t)  = e \vect{B}(\vect{x},t)/(mc)$ with the Lorentz factor 
$\gamma(t) = 1/ \sqrt{1-(v(t)/c)^2}$ , $c$ the speed of light in vacuum
and $\vect{u} (t) = \gamma \vect{v} (t)/c$. The local electric field $\vect{E}(\vect{x},t)$ on the $i$-th side of the shock 
is calculated within the ideal infinitely conductive MHD approximation in the shock frame: 
${\cal E}(\vect{x},t) = e \vect{E}(\vect{x},t)/(mc) = - (\vect{U}_i)/c \times \vect{\Omega} (\vect{x},t) $.
We calculate the trajectory of the particle in a magnetic field
as a solution of Eq.(\ref{lorentz})
where $t$ is the time in the rest frame of the plasma,
coincident with time in the shock frame because $U_1 \ll c$.
The quantity $\vect{\Omega}(\vect{x}, t)$ in Eq.(\ref{lorentz}) is given by
$\vect{\Omega}(\vect{x}, t) = \vect{\Omega}_0+\delta \vect{\Omega}(\vect{x}, t)$
where $\vect{\Omega}_0 \equiv (e/mc\gamma)
\vect{B}_0$, in terms of the background magnetic field $\vect{B}_0$ constant and statistically uniform,
and $\delta \vect{\Omega}(\vect{x}, t)$ is the turbulent component varying in the three-dimensional space; 
the time variation is accounted for by the frozen-in condition.

We use the profile of the bulk velocity specified in Sect.\ref{numerical}, namely $U_x(x) = U_1$ for $x<0$, $U_x(x) = U_2$ for $x>0$. The magnetic field embedded in the plasma is solution of the induction equation 
in the ideal MHD approximation, advected along the flow \citep{jg07,gj09}:
\begin{eqnarray}
B_x (x,y,z,t) & = & B_x (x_0,y,z,t_0) \nonumber\\ 
B_y (x,y,z,t) & = & B_y (x_0,y,z,t_0) U_1/U(x) \nonumber\\ 
B_z (x,y,z,t) & = & B_z (x_0,y,z,t_0) U_1/U(x) 
\label{B_ind}
\end{eqnarray}
where $x_0$ and $t_0$ are related by the characteristics equation: $t - t_0 = \int_{x_0} ^x dx' / U(x')$. For convenience, we choose $x_0$ in the far upstream of the shock and use the characteristic equation to find $t_0$. If ${\bf B}$ is known at $(x_0, y, z, t_0)$, using \ref{B_ind} the field can be calculated at every point where ${\bf B}$ is advected: at any time the magnetic field at particle position depends 
on the previous history of the field line carried by the flow.

The fluctuating field comprises of transverse waves at every point of physical space occupied by the particle, 
each with a random amplitude, phase, orientation and polarization \citep{gj99,fg12}. 
We use a three-dimensional isotropic, or Kolmogorov, power spectrum in the upstream plasma. 
The coherence scale $L_c$ of the turbulence is chosen such that $r_g^0/L_c \ll1$. 
The discrete modes are logarithmically equispaced. 
As in \citet{fg12} the randomization is performed over the initial particle velocity orientation 
in the local plasma frame and over the turbulence realization, shuffled every 50 particles.
Including the fluctuating field statistics ensures a meaningful comparison 
with a theoretically computed turbulence power spectrum,
which is by definition an average over an ensemble of field realizations. 

We inject a large number of particles on the plane $x = 0.5 r_g^0$,
with randomly oriented velocity, but fixed plasma frame initial kinetic energy. 
We integrate the Eq. (\ref{lorentz}) by using a time-step adjustable 
Burlisch-Stoer method \citep{p86}. The integration of the equation of motion is performed in the shock frame
which allows to calculate explicitly at every time step the ideal MHD electric field specified above.
This approach is well suited to investigate the particle transport in turbulence. 
A different approach of computing different realizations of the turbulent magnetic field 
in every cell of an Eulerian grid is much more time-consuming. 
The latter approach would also require adapting 
the lattice spacing in order to maintain the same space resolution in physical space.
Also, we did not implement the ``particle-splitting'' method as the momentum boundary 
is relatively small ($p < p_b = 500 p_0$).

In the steady state the boundary conditions are to be given in position and momentum \citep{g05}:
the trajectory of each particle is calculated until it reaches a far downstream return spatial boundary at 
$x_b = 2.5 \times 10^4 U_1/ \Omega_0$, or until the momentum $p$ in the plasma frame 
becomes greater than the upper boundary $p_b = 500 p_0$. 
If $p > p_b$, the particle is removed from the system. 
If $x > x_b$, we use the probability return algorithm 
\citep{ebj96,g05}: if the particle passes the test, it is reinitialised at the boundary plane $x = x_b$
with the same kinetic energy and same $y,z$ coordinates held at the last time-step 
before the boundary crossing with an isotropic pitch-angle distribution in the local downstream  plasma frame.
The re-injection mimics the returning flux of energetic particles at the location $x_b$ from the far downstream \citep{ebj96}. In fact, if we replace the downstream boundary at $x = x_b$ with an open boundary, at a location close to the boundary $x < x_b$ we find the expected isotropic distribution. 
In these numerical experiments we used $x_b > L_c$. 
This is consistent with the fact that the coordinates $(y_0, z_0)$ 
in the injection plane are spread over a scale greater than $L_c$: the simulation samples various regions of the shock both in the shock plane and in the perpendicular direction.

\begin{figure}
\includegraphics[width=9.5cm]{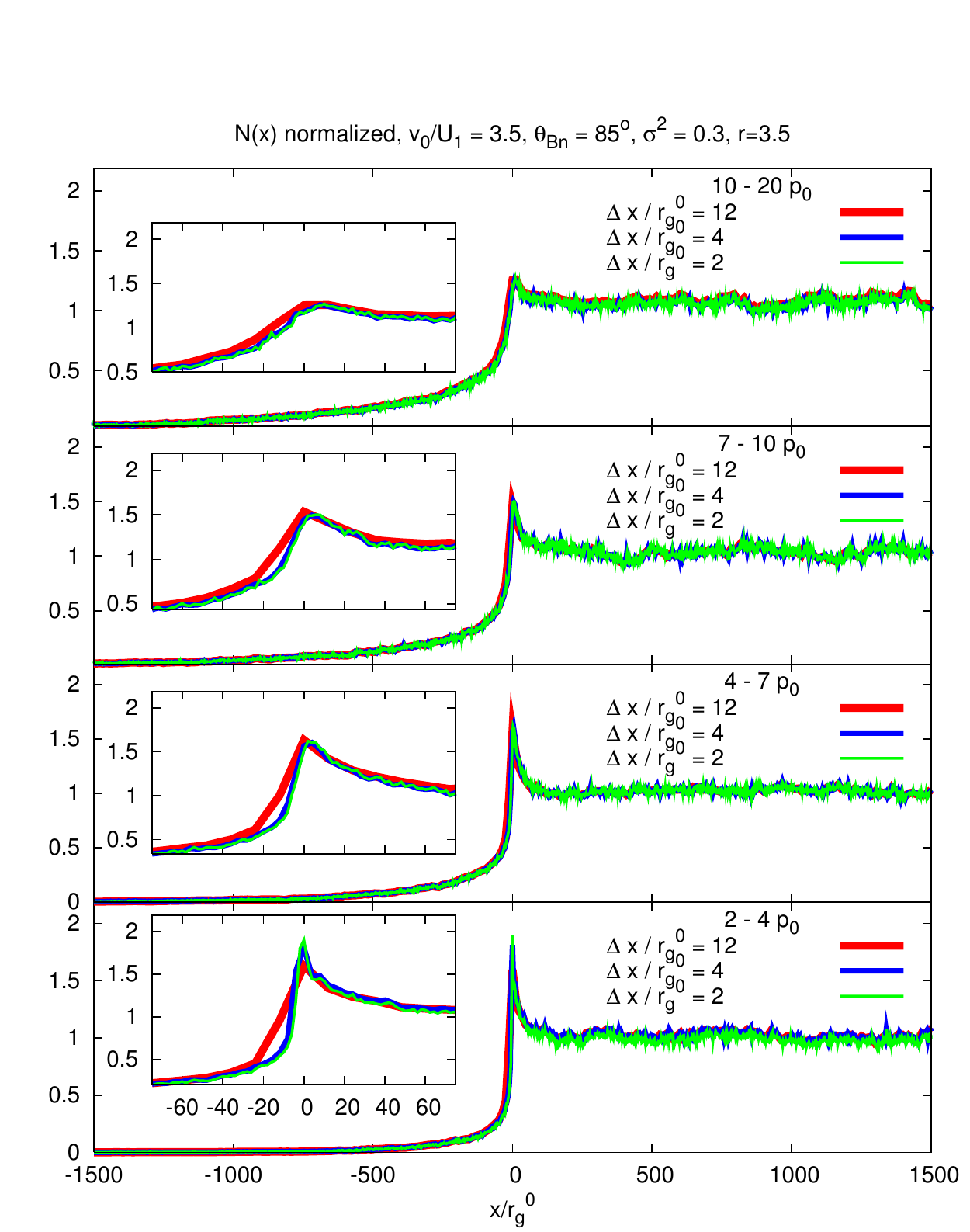}
\caption{Numerical convergence for various sampling interval $\Delta x/r_g^0 = 2, 4, 12$ in all momentum bands considered in this paper. The small panels on the left zoom into the shock in the same units.  \label{num_conv}}
\end{figure}

The steady-state density profiles comprise pseudo-particles: with a fixed cadence in time ($\Delta t \Omega_0 = 5$) and given a particle momentum range in the local plasma frame, we sample the position of each particle in spatial bins with uniform thickness along the $x$-axis from $-x_b$ to $x_b$ until each particle is removed from the system (the resulting spikes in the density profiles are verified to be roughly independent on the thickness of the bins). The density profile within the given momentum range is found by summing up in every spatial bin the total number of pseudo-particles within the given momentum range collected at every time. The interval $\Delta t \Omega_0$ is conveniently chosen not too large to track the gyro motion of the particles around the shock and not too small to reduce computational time. Likewise, the spectrum is produced by sampling the momentum of each particle in momentum bins with uniform thickness in logarithmic scale within $[0.9 p_0 : p_b]$ with cadence $\Delta t \Omega_0 = 5$. Figure \ref{num_conv} shows the numerical convergence of the spatial profiles for various spatial sampling interval $\Delta x/r_g^0$. 

\bsp

\label{lastpage}

\end{document}